\documentclass[conference]{IEEEtran}
\IEEEoverridecommandlockouts
\usepackage{verbatim}
\usepackage{enumitem}
\usepackage{algorithm}
\usepackage{algorithmic}
\usepackage[dvipsnames]{xcolor}
\usepackage{graphicx, color}
\usepackage{float}
\usepackage{subcaption}
\usepackage{amsmath}
\usepackage{amsfonts}
\usepackage{multirow}
\usepackage{colortbl}
\usepackage{placeins}
\usepackage{listings}
\usepackage{wrapfig}
\usepackage{hyperref}
\usepackage{tikz}
\usetikzlibrary{shapes}
\usetikzlibrary{calc}
\renewcommand{\algorithmicrequire}{\textbf{Input: }}

\newcommand{\ignore}[1]{}

\newcommand{\DNN}{\mathcal{N}}
\newcommand{\Spec}{\mathcal{S}}
\newcommand{\C}{\mathcal{C}}
\newcommand{\R}{\mathbb{R}}

\def\BibTeX{{\rm B\kern-.05em{\sc i\kern-.025emb}\kern-.08emT\kern-.1667em\lower.7ex\hbox{E}\kern-.125emX}}
    
\begin{document}
\title{Distribution-Aware Testing of Neural Networks Using Generative Models}

\author{\IEEEauthorblockN{Swaroopa Dola}
\IEEEauthorblockA{\textit{Department of Computer Engineering} \\
\textit{University of Virginia}\\
Charlottesville, USA \\
sd4tx@virginia.edu}
\and
\IEEEauthorblockN{Matthew B. Dwyer}
\IEEEauthorblockA{\textit{Department of Computer Science} \\
\textit{University of Virginia}\\
Charlottesville, USA \\
matthewbdwyer@virginia.edu}
\and
\IEEEauthorblockN{Mary Lou Soffa}
\IEEEauthorblockA{\textit{Department of Computer Science} \\
\textit{University of Virginia}\\
Charlottesville, USA \\
soffa@virginia.edu}
}

\maketitle

\begin{abstract}
The reliability of
software that has a Deep Neural Network (DNN) as a component is urgently important today given the increasing number of critical applications being deployed with DNNs.  The need for reliability raises a need for rigorous testing of the safety and trustworthiness of these systems. In the last few years, there  have been a number of research efforts focused on testing DNNs. However the test generation techniques proposed so far lack a check to determine whether the test inputs they are generating are valid, and thus invalid inputs are produced.  To illustrate this situation, we explored three recent DNN testing techniques. Using deep generative model based input validation, we show that all the three techniques generate significant number of invalid test inputs. We further analyzed the test coverage achieved by the test inputs generated by the DNN testing techniques and showed how invalid test inputs can falsely inflate test coverage metrics. 

To overcome the inclusion of invalid inputs in testing, we propose a technique to incorporate the valid input space of the DNN model under test in the test generation process. Our technique uses a deep generative model-based algorithm to generate only valid inputs. Results of our  empirical studies show that our technique is effective in eliminating invalid tests and boosting the number of valid test inputs generated.
\end{abstract}


\section{Introduction}
Deep Neural Networks (DNN) components are increasingly being deployed in mission and safety critical systems, 
e.g., \cite{end2endautonomous2017,pendleton2017perception,TrailNet:IROS2017,loquercio-etal:RAL:2018:dronet}. 
Similar to traditional \textit{programmed} software components, these
\textit{learned} DNN components require significant testing to 
ensure that they are reliable and thus fit for deployment.

Yet DNNs differ from programmed software components in a variety of ways.
(1) They generally do not have well-defined specifications and instead rely on 
a set of examples that represent intended component behavior.
(2) These examples are used to train the parameters of a fixed
implementation architecture resulting in implementation behavior 
encoded as values of the learned parameters.
(3) The training process continues until the learned function is 
an accurate approximation of the intended behavior.
Finally, (4) the accuracy of the learned function 
is intended to generalize to the set of valid inputs comprised of
the data distribution of which the training examples are representative.

The above characteristics of DNNs present challenges for applying existing software testing
methods to DNNs.  For example, the lack of specifications makes it most challenging
to develop a rich test oracle, as well as the fact that parameter values encode behavior
which renders traditional structural code coverage ineffective.
The growing body of research on DNN testing has begun to address some of 
these characteristics.
While structural code coverage metrics are ineffective for DNNs,
methods that cover combinations of computed DNN neuron values
have been developed to assess and drive DNN testing~\cite{pei2017deepxplore,ma2018deepgauge,sun2018testingmcdc}. 
Also, variations of metamorphic testing have been developed to check critical
continuity properties across the learned function 
approximations helping to fill the oracle gap~\cite{DBLP:journals/jss/XieHMKXC11,tian2018deeptest,huang2020survey}.
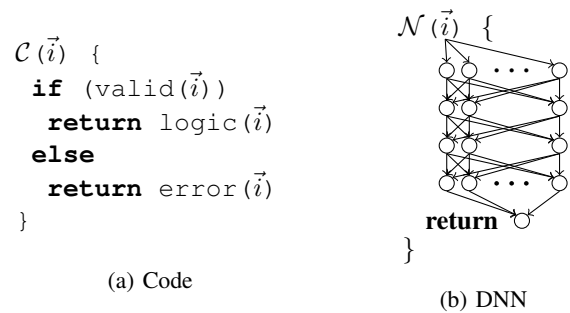
\begin{figure}[b]
\begin{subfigure}{.5\columnwidth}
  \centering
\begin{lstlisting}
    (*$\C$*)((*$\vec{i}$*)) {
      if (valid((*$\vec{i}$*)))
        return logic((*$\vec{i}$*))
      else
        return error((*$\vec{i}$*))
    }
\end{lstlisting}%
  \caption{Code}
  \label{fig:code}
\end{subfigure}%
\begin{subfigure}{.5\columnwidth}
  \centering
      \begin{tikzpicture}

\node[name=start] at (0,0.6) { $\mathcal{N}$\texttt{(}$\vec{i}$\texttt{) \{} };

\node[draw,circle,inner sep=0pt,minimum size=2mm,name=i1] at (0,0) {};
\node[draw,circle,inner sep=0pt,minimum size=2mm,name=i2] at (0.3,0) {};
\node[] at (0.9,0) {\Large \ldots};
\node[draw,circle,inner sep=0pt,minimum size=2mm,name=in] at (1.5,0) {};

\node[draw,circle,inner sep=0pt,minimum size=2mm,name=h1] at (0,-0.5) {};
\node[draw,circle,inner sep=0pt,minimum size=2mm,name=h2] at (0.3,-0.5) {};
\node[] at (0.9,-1.5) {\Large \ldots};
\node[draw,circle,inner sep=0pt,minimum size=2mm,name=hn] at (1.5,-0.5) {};

\node[draw,circle,inner sep=0pt,minimum size=2mm,name=O1] at (0,-1.5) {};
\node[draw,circle,inner sep=0pt,minimum size=2mm,name=O2] at (0.3,-1.5) {};
\node[] at (0.9,-1.5) {\Large \ldots};
\node[draw,circle,inner sep=0pt,minimum size=2mm,name=On] at (1.5,-1.5) {};

\node[draw,circle,inner sep=0pt,minimum size=2mm,name=p1] at (0,-1.0) {};
\node[draw,circle,inner sep=0pt,minimum size=2mm,name=p2] at (0.3,-1.0) {};
\node[draw,circle,inner sep=0pt,minimum size=2mm,name=pn] at (1.5,-1.0) {};

\node[name=return] at (0.2,-2.0) {\texttt{\bf return}};
\node[draw,circle,inner sep=0pt,minimum size=2mm,name=O] at (1.0,-2.0) {};
\node[name=close] at (-0.5,-2.4) {\texttt{\}}};

\draw[->] ($(start.south)+(-0.2mm,1.5mm)$) -- (i1.north);
\draw[->] ($(start.south)+(-0.2mm,1.5mm)$) -- (i2.north);
\draw[->] ($(start.south)+(-0.2mm,1.5mm)$) -- (in.north west);

\draw[->] (i1.south) -- (h1.north);
\draw[->] (i1.south) -- (h2.north);
\draw[->] (i1.south) -- (hn.north west);

\draw[->] (i2.south) -- (h1.north);
\draw[->] (i2.south) -- (h2.north);
\draw[->] (i2.south) -- (hn.north west);

\draw[->] (in.south) -- (h1.north east);
\draw[->] (in.south) -- (h2.north east);
\draw[->] (in.south) -- (hn.north);

\draw[->] (h1.south) -- (p1.north);
\draw[->] (h1.south) -- (p2.north);
\draw[->] (h1.south) -- (pn.north west);

\draw[->] (h2.south) -- (p1.north);
\draw[->] (h2.south) -- (p2.north);
\draw[->] (h2.south) -- (pn.north west);

\draw[->] (hn.south) -- (p1.north east);
\draw[->] (hn.south) -- (p2.north east);
\draw[->] (hn.south) -- (pn.north);

\draw[->] (p1.south) -- (O1.north);
\draw[->] (p1.south) -- (O2.north);
\draw[->] (p1.south) -- (On.north west);

\draw[->] (p2.south) -- (O1.north);
\draw[->] (p2.south) -- (O2.north);
\draw[->] (p2.south) -- (On.north west);

\draw[->] (pn.south) -- (O1.north east);
\draw[->] (pn.south) -- (O2.north east);
\draw[->] (pn.south) -- (On.north);

\draw[->] (O1.south) -- (O.north west);
\draw[->] (O2.south) -- (O.north west);
\draw[->] (On.south) -- (O.north east);

\end{tikzpicture}
  \caption{DNN}
  \label{fig:dnn}
\end{subfigure}
    \caption{Structure of code and DNN components $\C$ and $\DNN$.}
    \label{fig:structure}
\end{figure}
In this paper, we focus on the challenges that DNN generalization presents
to testing, and in particular how current DNN testing techniques treat valid and invalid inputs. 
To understand these challenges, consider the implementation of a traditional software component
$\C$, which is developed to meet a specification 
$\Spec : \R^{n} \rightarrow \R^{m} \cup e$, 
where $e$ denotes the error behavior intended
for \textit{invalid inputs}.
In this setting, the input domain $\R^{n}$ is partitioned into 
valid inputs, $V$, and invalid inputs, $\overline{V} = \R^{n} - V$, which should yield $e$.

\begin{figure*}[t]
    \centering
    {\tt
    $V$~~~~0000000100000000000000001\textcolor{ForestGreen}{1}00100011111111111111111111~~~~~~~0.462~~~~~~~~~\\
    $\overline{V}$~~~~0000000100000000000000001000100011111111111111111111~~~~~~~0.442 (0.462)\\
    \vspace{-1mm}\noindent\hrulefill\\
    $V$~~~~1101110101101000110001101000110011111111111111111111~~~~~~~0.692~~~~~~~~~\\
    $\overline{V}$~~~~11\textcolor{Red}{1}111\textcolor{Red}{1}1000010\textcolor{Red}{1}00100011010\textcolor{Red}{1}0010\textcolor{Red}{1}11111111111111011011~~~~~~~0.673 (0.808)\\
    }
    \caption{Cumulative neuron coverage of LeNet-1 on the first 100 valid and invalid inputs generated by DLFuzz (top) and DeepXplore (bottom); coverage vectors (left) and ratios (right) for each set are shown along with the cumulative ratio (in parentheses)}
    \label{fig:coveragevectors}
\end{figure*}

The testing of $\C$ selects a test set $T \subset \R^{n}$ and assesses
whether $\forall t \in T : \C(t) = \Spec(t)$.
As sketched in Fig.~\ref{fig:code}, typically $\C$ is comprised of \textit{input validation},
which determines if an input value lies in $\overline{V}$ and then executes either
\textit{functional logic} which realizes the behavior of $\Spec$ on $V$, or 
\textit{error processing} for invalid data.
Developers have come to rely on the several intuitions about such software.
First, input validation logic is distinct from functional logic, demanding testing 
approaches that exploit its 
properties~\cite{hayes2006input,li2010perturbation,liu2009covering,taneja2010mitv}
to effectively support 
it~\cite{sinha2000analysis,zhang2014amplifying,goffi2016automatic}.
Second, \textit{test suites that achieve higher coverage are better} in that they
exercise more of the validation, functional, and error logic.

Now, consider a DNN, $\DNN : \R^{n} \rightarrow \R^{m}$, which is trained
to accurately approximate the, possibly unavailable, specification $\Spec$.
As sketched in Fig.~\ref{fig:dnn}, $\DNN$ is comprised of layers of neurons
that are cross-coupled by connections labeled with learned parameters.  
When the learned parameters for $\DNN$ are such that
$Pr(\DNN(i) = \Spec(i) \mid i \in V) \ge 1 - \epsilon$, for a desired error
$\epsilon$, the network is expected to generalize to the valid input distribution, $V$.
Even if $\DNN$ were trained to detect invalid data and respond appropriately, 
its structure does not force a distinction between input validation, functional logic,
or error processing.  
In practice, this distinction is uncommon and in this case $\DNN$ does not even have an analog for $e$ in its output domain.
Because of the lack of this distinction, whether an input lies in 
$V$ or $\overline{V}$, the computation performed by $\DNN$ overlaps 
to a large degree, e.g., 
common sets of neurons are activated.

\begin{figure}
    \centering
     \includegraphics[width=.1\textwidth]{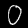}
    \includegraphics[width=.1\textwidth]{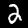}
    \includegraphics[width=.1\textwidth]{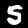}
    \medbreak
    \includegraphics[width=.1\textwidth]{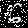}
    \includegraphics[width=.1\textwidth]{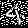}
    \includegraphics[width=.1\textwidth]{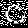}
    \medbreak
    \includegraphics[width=.1\textwidth]{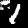}
    \includegraphics[width=.1\textwidth]{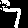}
    \includegraphics[width=.1\textwidth]{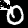}
    \break
    \caption{Valid tests vs Invalid tests. Top Row: Valid tests from MNIST training dataset. Middle Row: Invalid tests from DeepConcolic. Bottom Row: Invalid tests from DeepXplore}
    \label{fig:invalidimages}
\end{figure}

Not distinguishing between valid and invalid input can be problematic for DNN testing in at least three ways.
(1) Testing techniques that generate invalid inputs increase cost
with little value added for testing the functional logic of $\DNN$.
Fig.~\ref{fig:invalidimages} depicts valid test inputs and selected invalid test inputs from two
recently proposed DNN test generation techniques~\cite{pei2017deepxplore,sun2018concolic}.
As we show in \S\ref{sec:evaluation}, across a range of testing approaches for DNNs~\cite{guo2018dlfuzz,pei2017deepxplore,sun2018concolic}, on average 42\% of the generated tests are invalid and in the worst-case all generated tests by a given technique are invalid.
(2) When a test case fails developer time is required to triage the failure. With  high numbers of invalid test inputs,
developers may be forced to look through large numbers of test inputs, 
similar to those depicted in Fig.~\ref{fig:invalidimages}, to make judgements
about test validity.   The
high-rate of invalid inputs runs the risk that developers will avoid
the use of these techniques, thereby negating their purported value.
(3) Whereas for traditional software the coverage produced by invalid inputs
is confined to the validation and error logic, for DNNs an analogous separation 
of coverage is not guaranteed.  As depicted at the top of Fig.~\ref{fig:coveragevectors},
the cumulative coverage from valid and invalid test sets can be almost identical -- differing
by as few as 1 of 52 neurons.
Worse yet, as depicted in the bottom of Fig.~\ref{fig:coveragevectors}, invalid tests
can artificially boost coverage significantly beyond what is achieved by valid tests - from 0.692 to 0.808.
This increase in coverage suggests that, unlike for traditional software,
\textit{DNN test suites that  achieve higher coverage are not necessarily better}!

In this paper, we study the effects of DNN test generation techniques not distinguishing between valid and invalid data
and characterize the potential impact of the issues identified above.
Our approach is to leverage a growing body of research from the Machine Learning (ML) community that learns
models of the training distribution, $V$, from which the training data
is drawn~\cite{an2015variational, xu2018unsupervised, zenati2018efficient,  chalapathy2019deep}. 
While there are many such models, in this paper we employ the \textit{variational auto-encoder} (VAE) -- leaving the study of alternative models to future work.

Leveraging VAE models allows us to study techniques representative of
the current state of DNN testing research and to make two important observations.
First, we demonstrate that existing DNN testing techniques, such as
DeepXplore~\cite{pei2017deepxplore},
DLFuzz~\cite{guo2018dlfuzz}, and DeepConcolic~\cite{sun2018concolic},
produce large numbers of test cases with invalid inputs, which increases
test cost without a clear benefit.
Second, we demonstrate that existing DNN test coverage metrics, e.g.,
\cite{pei2017deepxplore,ma2018deepgauge}, 
are unable to distinguish valid and invalid test cases, which risks biasing
test suites toward including more invalid inputs in pursuit of higher coverage.

Building on these observations, we present a novel approach
that combines a VAE model with existing test generation techniques to
produce test cases with only valid inputs. More specifically, we 
formulate the joint optimization of probability density of valid inputs and 
the objective of existing DNN test generation techniques, and use gradient ascent
to generate valid tests. 
An experimental analysis on datasets used in the DNN testing 
literature~\cite{lecun1998mnist,netzer2011reading} shows the 
cost-effectiveness of the proposed approach.

The primary contributions of this work lie in:
(a) the identification of
limitations in existing DNN test generation and coverage criteria in their treatment of invalid input data;
(b) the development of a technique for incorporating an explicit model of the valid
input space of a DNN into test generation to address those limitations; and
(c) experimental evaluation that demonstrates the extent of the limitations and the effectiveness of our technique in mitigating them.

The remainder of this article is organized as follows.  
The following section, \S\ref{background}, 
describes the concepts that are used in this paper and related work. Our
approach is detailed in \S\ref{approach}.  Experimental strategy and
results are described in \S\ref{sec:evaluation}.  \S\ref{threats} discusses the threats to validity of our
study and \S\ref{conclusions} concludes.

\section{Background and Related Research}\label{background}
\subsection{Deep Neural Networks}
Deep Neural Networks (DNNs) are a class of Machine Learning models that can extract high level features from raw input. Similar to the human brain, DNNs contain a large number of inter-connected elements called neurons. DNNs have multiple layers, and each layer contains a number of neurons. A typical DNN consists of an input layer, one or more hidden layers followed by an output layer. Connections between neurons are called edges and their associated weights are referred to as the model parameters. A neuron receives its input as a weighted sum over outputs of neurons from the previous layer. The neuron then applies a non-linear activation function on this input to generate its output. Overall, a DNN is a mathematical function over the model parameters for transforming inputs into outputs. The model learns its parameters by training on known input data called the training data. The objective of DNN training is to learn the model parameters in order to make accurate predictions on unseen data during deployment.

\subsection{DNN testing techniques}
DNN testing is an active research area with a number of testing techniques developed to address the challenges of testing these systems \cite{zhang2020machine,huang2020survey} in terms of test coverage criteria, test generation and test oracles.

After training, DNN testing techniques use either natural inputs or adversarial inputs for testing. Adversarial inputs are test inputs that are generated by applying tiny perturbations on the original inputs, which cause the model to make false predictions~\cite{goodfellow2014explaining}. There is another line of research that focuses on generating adversarial examples for exposing vulnerabilities of DNN models~\cite{goodfellow2014explaining, kurakin2016adversarial, carlini2017towards} without addressing test adequacy. However our work differs by focusing on coverage guided DNN testing techniques from the software engineering literature. 

\subsubsection{Coverage Criteria}
In traditional software testing, coverage criteria are used to measure how thoroughly software is tested. Most practical coverage criteria e.g., \cite{stuctural_coverage}, use the structure of the software system to make this assessment, e.g., the percentage of statements or branch outcomes covered by a test suite.  Similar to structural software coverage criteria, coverage criteria for DNNs have been proposed by various research efforts, as follows.

Pei et al.~\cite{pei2017deepxplore} proposed neuron coverage (NC) as a test coverage criteria. For a given test suite, neuron coverage is measured as the ratio of the number of unique neurons whose output exceeds a specified threshold value to the total number of neurons present in the DNN. 

Ma et al.~\cite{ma2018deepgauge} proposed a range of coverage criteria including:  k-multisection neuron coverage (KMNC), neuron boundary coverage (NBC), and strong neuron activation coverage (SNAC). These coverage criteria can be used to  determine whether a test case falls in the major functional region or corner case region of a DNN. Activation traces of all neurons are captured for the training data and lower and upper bounds of activations are measured for each of the neurons. 

K-multisection coverage is calculated by dividing the interval between lower and upper bounds into k-bins and measuring the number of bins activated by the test inputs. For a test suite, k-multisection coverage is the ratio of the uniquely covered bins to the total number of bins in the model. 

Neuron activations above the upper bound or below the lower bound are considered to be in corner case regions. Neuron boundary coverage is measured as a ratio of the number of covered upper and lower corner case regions to the total number of corner case regions of the model. Strong neuron activation coverage is the ratio of the number of covered upper corner case regions to the total number of upper corner case regions in the DNN. Top-k neuron coverage and top-k neuron patterns are based on top hyper-activate neurons and their combinations.

Modified Condition/Decision Coverage variants for DNNs \cite{sun2018testingmcdc} 
are proposed by Sun et al~\cite{sun2018testingmcdc}. These metrics are based on 
sign and value change of a neuron's activation to capture the causal changes 
in the test inputs. Ma et al.~\cite{ma2018combinatorial} proposed combinatorial 
test coverage to measure the combinations of neuron activations and deactivations 
covered by a test suite.

In our work, we focus on the NC, KMNC, NBC, and SNAC criteria and we show that these metrics cannot differentiate between valid and invalid test inputs generated by existing DNN test generation techniques. We leave the analysis for other coverage metrics for future work. 

\subsubsection{DNN test generation}
Research on DNN test generation is largely inspired by traditional software testing techniques such as metamorphic testing, fuzz based testing and symbolic execution. Below, we discuss the state of DNN test generation research.

DeepXplore~\cite{pei2017deepxplore} is a white-box differential test generation technique that uses domain specific constraints on inputs. This technique requires multiple DNN models trained on the same dataset as cross referencing oracles. The objective of DeepXplore is a joint optimization of neuron coverage and differences in the predictions of DNN models. Maximizing the objective generates tests that achieve high neuron coverage while simultaneously achieving erroneous predictions by the DNN model. DeepXplore uses gradient ascent to solve the joint optimization. DeepTest~\cite{tian2018deeptest} is another testing technique that generates test inputs by applying domain specific constraints on seed inputs. The major focus of DeepTest is to generate test inputs for testing autonomous vehicles. It uses greedy search driven by neuron coverage criteria.

Fuzzing is another traditional software testing technique that has been adapted for DNN test generation including DLFuzz~\cite{guo2018dlfuzz}, and TensorFuzz~\cite{odena2019tensorfuzz}. DLFuzz is an adversarial input test generation technique. It uses neuron coverage driven test generation similar to DeepXplore. However unlike DeepXplore, it does not require multiple DNN models. It also uses a constraint to keep the newly generated test inputs close to the original inputs. TensorFuzz is a coverage guided testing method for finding numerical issues in trained neural networks and disagreements between neural networks and their quantized versions.

DeepConcolic~\cite{sun2018concolic} uses the concolic testing approach for generating adversarial test inputs for DNN testing. Concolic execution is a coverage-guided testing technique that combines symbolic execution and path information from concrete execution for generating tests satisfying a coverage criteria. DeepConcolic supports neuron coverage and MC/DC variants for DNNs.

None of these DNN testing techniques check whether the test inputs they are generating follow the training distribution. They generate a significant number of invalid inputs that are outside the model's training distribution as shown in our evaluation section~\ref{sec:evaluation}. 

\subsection{Out-of-Distribution Input Detection}
Out-of-distribution input detection (OOD), also referred to as outlier or anomaly detection, is a well-studied problem in ML field\cite{an2015variational, xu2018unsupervised, zenati2018efficient, hendrycks2018deep}. A recent survey~\cite{chalapathy2019deep} describes the state of deep learning based outlier detection research and classifies deep learning based outlier detection techniques into supervised, semi-supervised, unsupervised categories. Unsupervised models are preferred as labeling is expensive. We use an unsupervised generative model based approach for our work. 

A generative model learns the distribution of the data and can predict how likely a test input is with respect to training distribution. This prediction can be used to identify invalid test inputs. A DNN classifier learns the conditional distribution of target variables with respect to observable variables. Even though such a classifier has high accuracy on data sampled from the training distribution, its accuracy on samples outside the training distribution cannot be guaranteed~\cite{nguyen2015deep}. By training a  generative model with the same data, its density predictions can be used to reject inputs with low densities. When a test input has low density it implies that the DNN classifier did not have enough samples around test input region in the training dataset. 

Examples of generative models are autoencoders, variational autoencoders~\cite{kingma2013auto}, generative adversarial networks (GAN)~\cite{goodfellow2014generative}, and autoregressive models such as PixelCNN~\cite{van2016conditional} and PixelCNN++~\cite{salimans2017pixelcnn++}. We primarily use the variational autoencoder based out-of-distribution detection technique in our work.  Also, we repeat our experiments to identify invalid inputs generated by test generation techniques using a PixelCNN++ based validation approach. The study is described in section~\ref{results} to show how sensitive invalid input identification is with respect to the out-of-distribution detection mechanism used.

\subsection{Variational Autoencoder}
A variational autoencoder is a generative model that represents latent space as a probability distribution. It has an encoder, code layer and a decoder~\cite{kingma2013auto}. The encoder is responsible for mapping inputs to a lower dimensional latent space, and the decoder generates new inputs by sampling from the latent space. Latent space is modeled by a code layer, and it is generated from a prior distribution, e.g., a Normal Gaussian distribution. The encoder's objective is to learn the posterior distribution and decoder's objective is to learn the likelihood of the original input reconstructed by the decoder. A VAE model is trained by minimizing the difference between posterior and latent prior distributions and maximizing the likelihood estimation of the input. A trained VAE model will generate high probability density estimates for data belonging to the training data distribution when compared to out-of-distribution inputs. This key insight is used for validating test inputs generated by DNN test generation techniques in our research.

\section{Approach}\label{approach}
In this section, we describe our approach to (1) identifying limitations of existing DNN test generation techniques, and (2) generating valid test inputs for testing DNNs.

\subsection{Analysis of Existing DNN Test Generation Techniques}\label{threshold}
\begin{figure}
  \centering
  \includegraphics[width=\linewidth]{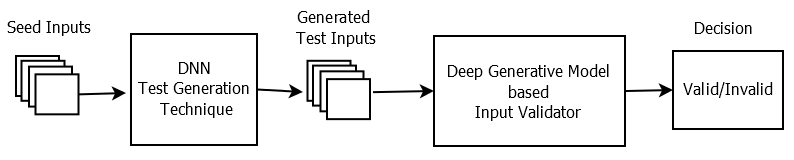}
  \caption{Technique for identifying invalid test inputs}
\label{validation}
\end{figure}

The methodology for analysing test inputs generated by existing test generation techniques is depicted in Fig.~\ref{validation}. DNN(s) under test and the deep generative model are trained on the same dataset. Test inputs generated by existing DNN test generation techniques for the DNN(s) under test are passed as inputs to the deep generative model which estimates their densities. These densities are used by the decision logic to classify inputs as valid or invalid.

For our experiments, we use a VAE for expressing the deep generative model logic, and in particular, the model proposed by An and Cho \cite{an2015variational} where the decoder of a VAE outputs distribution parameters for the samples generated by the encoder. The probability of generating the original test input from a latent variable is calculated using these distribution parameters. This probability is referred to as reconstruction probability. Valid inputs have higher reconstruction probability when compared to invalid inputs.

For a dataset under test, which we call the valid dataset, we identify another dataset which has a different distribution. The inputs from this dataset are considered as invalid inputs. Invalid dataset selection is guided by two factors: (1) the dataset should have same input dimensions as the valid dataset, and  (2) invalid and valid datasets should model disjoint data categories. 

After identifying an invalid dataset, we compute the reconstruction probability threshold for identifying invalid inputs. Reconstruction probabilities are calculated for inputs from both valid and invalid datasets. We generate a range of thresholds from the combined reconstruction probability values of valid and invalid inputs. We compute the F-measure, which is a measure of a test's accuracy, for these threshold values. The F-measure is the harmonic mean of precision and recall. A good F-measure balances precision and recall and results in a fewer number of both false positives and false negatives. In our  case, this means  fewer  valid inputs are falsely classified as invalid and  fewer invalid inputs  are falsely classified as valid. The threshold value with the highest F-measure is selected for our experiments. When classifying test inputs generated by DNN test generation techniques, test inputs with reconstruction probability less than the selected threshold are classified as invalid by the VAE classifier.

We measure the percentage of invalid inputs generated by multiple test generation techniques and the coverage of both valid and invalid tests. The results of the experiments are used to answer the research questions related to the limitations of existing techniques presented in Section~\ref{sec:evaluation}.

\subsection{Our Test Generation Technique}
\begin{figure}
  \centering
  \includegraphics[width=\linewidth]{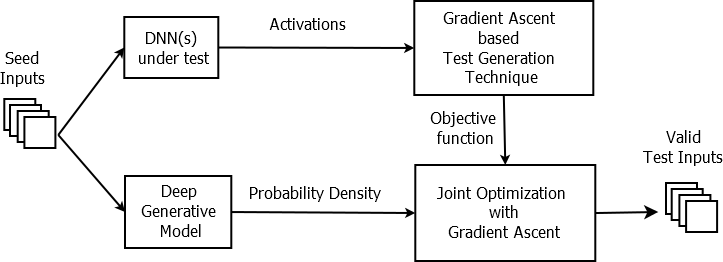}
  \caption{Technique for generating valid test inputs}
\label{generator}
\end{figure}

We present a technique to generate valid test inputs in this section. Our workflow is described in Fig.~\ref{generator}. Our approach leverages existing gradient ascent based test generation technique's objective formulation. The objective of existing test generation techniques is modeled to increase test coverage and produce inputs that cause the model to make incorrect predictions. We augment this objective with probability density estimated by a generative model. Gradient ascent is used to solve the joint optimization. Maximizing the joint optimization will result in inputs that follow the distribution of the training data of the DNN under test along with satisfying objective of the baseline testing technique.

We provide a detailed description of our test generation algorithm using a VAE as the generative model in Algorithm~\ref{alg1}. The decoder of the VAE outputs the distribution parameters ($\mu_{\text{\^x}^{}}$, $\sigma_{\text{\^x}^{}}$) for the samples generated by the encoder as per the OOD detection algorithm proposed in \cite{an2015variational}. The algorithm requires a DNN under test, an objective function of a baseline gradient ascent based test generation technique $obj1$, a probabilistic encoder and decoder as inputs and produces both a test suite of valid inputs and their test coverage as output. For every input of the seed set, the probabilistic encoder generates parameters in latent space as shown in line 4 of the Algorithm~\ref{alg1}. In lines 5-7, a sample from the latent space is used by the decoder to calculate the reconstruction probability of the input. The objective is modeled as a weighted sum of $obj1$ and reconstruction probability in line 8. Lines 9-11 show the gradient ascent. The gradient is calculated for the objective and at this stage, domain constraints, if any, are applied to the gradient and a new test input is generated. In lines 12-13, the generated test is tested for validity. If this test input causes the model to mispredict and has a reconstruction probability higher than the threshold, then on lines 14-15 the coverage is updated and input is added to the generated test suite.  The procedure continues until all seeds are processed. We evaluate this technique using DeepXplore as a baseline test generation technique in Section~\ref{sec:evaluation}.

\begin{algorithm}
\caption{Valid test input generation using VAE}

\algorithmicrequire 

X $\leftarrow$ Seed inputs \\
DNN $\leftarrow$ DNN under test \\
obj1 $\leftarrow$ Objective function of test generation technique\\
s $\leftarrow$ Step size for gradient ascent \\
max\_iterations $\leftarrow$ maximum iterations for gradient ascent \\
$f_\theta^{}$, $g_\phi^{}$ $\leftarrow$ Trained probabilistic encoder and decoder\\
$\lambda$ $\leftarrow$ hyperparameter for balancing two goals \\
$\alpha$ $\leftarrow$ Reconstruction probability threshold \\
\textbf{Output:} Set of test inputs, coverage

\begin{algorithmic}[1]
\STATE gen\_test = \{\}
\FOR{x in X}
    \FOR{i=1 \TO max\_iterations}
        \STATE $\mu_z^{}$, $\sigma_z^{}$ = $f_\theta^{}(z|x)$
        \STATE draw sample from z $\sim \mathcal{N}$($\mu_z^{}$, $\sigma_z^{}$)
         \STATE $\mu_{\text{\^x}^{}}$, $\sigma_{\text{\^x}^{}}$ = $g_\phi^{}(x|z^{})$  
        \STATE obj2 = $p_\theta^{}(x|\mu_\text{\^x}^{}$, $\sigma_\text{\^x}^{}$)
    
        \STATE obj = obj1 $+$ $\lambda$ $\times$ obj2
        \STATE gradient = ${\partial obj}/{\partial x}$
        \STATE gradient = Constraints(gradient)
        \STATE x = x + s $\times$ gradient
        \STATE p = Reconstruction\_Probability(x, $f_\theta^{}$, $g_\phi^{}$)
        \IF{Counter\_Example(DNN, x) $\AND$ p $\ge \alpha$}
        \STATE gen\_test.add(x)
        \STATE update coverage
        \STATE break
        \ENDIF
    \ENDFOR
\ENDFOR
\end{algorithmic}
\label{alg1}
\end{algorithm}

\section{Evaluation}\label{sec:evaluation}
The design and evaluation of experiments for studying existing techniques and demonstrating effectiveness of our approach are described in this section. We answer the following research questions:
\begin{description}[align=left]
\item[RQ1:] Do existing test generation techniques produce invalid inputs?
\item[RQ2:] Existing test generation techniques are guided by test coverage criteria. How do invalid inputs affect test coverage metrics? 
\item[RQ3:] VAE based input validation can be incorporated into test generation techniques. How effective is this technique in generating valid inputs and what is the overhead?
\item[RQ4:] Is the determination of invalid inputs sensitive to the generative model used?
\end{description}
\subsection{Evaluation Setup}
All experiments are conducted on servers with one Intel(R) Xeon(R) CPU E5-2620 v4 2.10GHz processor with 32 cores, 62GB of memory, and 4 NVIDIA TITAN Xp GPUs.  The software that supports our evaluation as well as all of the data described below is available at \url{https://github.com/swa112003/DistributionAwareDNNTesting}.

\subsubsection{Test Generation Frameworks}
We study three state of the art test generation techniques: DeepXplore~\cite{pei2017deepxplore}, DLFuzz~\cite{guo2018dlfuzz}, and DeepConcolic~\cite{sun2018concolic} to demonstrate the limitations of existing techniques in terms of generating valid test inputs and satisfying test coverage criteria. The choice of these frameworks is guided by the categorization of test input generation techniques presented in a recent survey \cite{zhang2020machine} and the availability of open source code. The survey categorizes test generation frameworks into three algorithmic families; we choose one technique from each family. DeepXplore is selected from domain-specific test input synthesis, DLFuzz from fuzz and search based test input generation and DeepConcolic from symbolic execution based test input generation categories.
\begin{table}
\centering
\begin{tabular}{ |l|l|l|l|l|l|l| } 
 \hline
 \textbf{Dataset} & \textbf{Name} & \multicolumn{2}{c|}{\textbf{Architecture}} & \textbf{Accuracy} \\ 
  &  & \textbf{Source} & \textbf{\#l:\#n:\#p}  &  \\ 
 \hline
 MNIST & \begin{tabular}{@{}l@{}l@{}}
           MNI-1 \\ MNI-2 \\ MNI-3 \\ MNI-4 \\ \end{tabular} & 
         
         \begin{tabular}{@{}l@{}l@{}}
           LeNet-1~\cite{lecun1998gradient} \\ LeNet-4~\cite{lecun1998gradient} \\ LeNet-5~\cite{lecun1998gradient} \\ Custom~\cite{sun2018concolic} \end{tabular} &

          \begin{tabular}{@{}l@{}l@{}} 3:52:7206 \\ 4:148:69362 \\ 5:268:107786 \\ 7:1300:312202 \end{tabular} &
          
         \begin{tabular}{@{}l@{}l@{}} 98.66\% \\ 99.03\% \\ 99.08\% \\ 99.03\% \end{tabular} \\
 \hline
 SVHN & \begin{tabular}{@{}l@{}l@{}}
                   SVH-1 \\ SVH-2 \\ SVH-3 \\ SVH-4 \\ \end{tabular} & 

                     \begin{tabular}{@{}l@{}l@{}}
                   ALL-CNN-A~\cite{springenberg2014striving} \\ ALL-CNN-B~\cite{springenberg2014striving} \\ ALL-CNN-C~\cite{springenberg2014striving} \\ 
                   VGG19~\cite{simonyan2014vggnet} \\ \end{tabular} & 

                \begin{tabular}{@{}l@{}l@{}} 7:2248:1.2M \\ 9:2824:1.3M \\ 9:2824:1.3M \\ 19:28884:38M \end{tabular} &
                \begin{tabular}{@{}l@{}l@{}} 96\% \\ 95.67\% \\ 95.98\% \\ 94.69\% \end{tabular} \\
 \hline
\end{tabular}
\caption{Models used in our studies with number of layers (\#l), neurons (\#n), parameters (\#p), and test accuracy; ``M'' denotes millions of parameters.}
\label{t1}
\end{table}
\subsubsection{Test Coverage Criteria}
DeepXplore and DLFuzz use neuron coverage~\cite{pei2017deepxplore} as the test adequacy criteria whereas DeepConcolic can be used with neuron coverage~\cite{pei2017deepxplore}, neuron boundary coverage~\cite{ma2018deepgauge} and MC/DC coverage criteria for DNNs~\cite{sun2018testingmcdc}. We use neuron coverage as the test adequacy criteria for generating tests using all three frameworks. Resulting test inputs from test generation are analyzed using neuron coverage and extended neuron coverage metrics, i.e, k-multisection neuron coverage, neuron boundary coverage and strong neuron activation coverage. We leave the remaining coverage criteria discussed in these works~\cite{ma2018deepgauge,sun2018testingmcdc} for future study.
\subsubsection{Datasets and DNN Models}
We use two popular datasets MNIST~\cite{lecun1998mnist} and SVHN~\cite{netzer2011reading} for the experiments. Generative models can assign higher densities to datasets whose distributions are different from their training datasets in some cases\cite{nalisnick2018deep}. For example, a VAE trained on CIFAR10~\cite{krizhevsky2009learning} can assign higher densities to inputs from SVHN dataset. When such a model is used for invalid input identification, it might result in high densities being assigned to invalid inputs which will result in false negatives. Also selecting the threshold density for deciding invalid inputs becomes challenging in such scenarios. This problem is actively being addressed by ML research community\cite{ren2019likelihood}. Generative models trained on MNIST and SVHN do not have this issue~\cite{nalisnick2018deep}, so we selected these two datasets for our research.

\textbf{MNIST} is a collection of grayscale images of handwritten digits with 60000 training images and 10000 test images. All three frameworks that we are studying support test generation for MNIST dataset. Similar to DeepXplore, we use LeNet-1, LeNet-4 and LeNet-5 networks from LeNet family~\cite{lecun1998gradient} and a custom architecture used in the DeepConcolic work~\cite{sun2018concolic} for MNIST classification. All the four models are convolutional networks with max-pooling layers and the number of layers ranging from 3 to 7.

\textbf{SVHN} contains color images of digits in natural scenes and the dataset has 73257 training images and 26032 test images. We implemented SVHN support for all three frameworks. We trained SVHN classification models with the ALL-CNN-A, ALL-CNN-B and ALL-CNN-C network architectures proposed in \cite{springenberg2014striving} and VGG19~\cite{simonyan2014vggnet} for our experiments. These models are convolutional networks with dropout and either global average pooling or max-pooling layers and the number of layers range from 7 to 19.
The models are summarized in Table~\ref{t1} where we report measures of their architecture and test accuracy.

\subsubsection{VAE Models}\label{eval_setup_vae}
For MNIST, we trained the VAE that outputs distribution parameters using the model architecture described in ~\cite{an2015variational}. The FashionMNIST dataset~\cite{xiao2017fashionmnist}, is similar to MNIST and contains 28x28 grey scale images. However the distribution is different from that of MNIST as FashionMNIST contains clothing images. We use the FashionMNIST as the invalid input space for calculating the reconstruction probability threshold. Since the VAE is not trained on FashionMNIST distribution and FashionMNIST clothing inputs are semantically unrelated to MNIST digit inputs, the VAE should output lower reconstruction probabilities for test inputs from the FashionMNIST dataset.

We experimented with different variations of the generator architecture used in~\cite{rosca2018distribution} for selecting a VAE network for the SVHN dataset. For each of the variants, the encoder is created by transposing the generator network as suggested in~\cite{rosca2018distribution}. The network that achieved the highest F-measure for identifying invalid inputs is selected for our experiments. CIFAR10~\cite{krizhevsky2009learning} is used as the invalid input dataset for calculating reconstruction probability threshold of VAE trained on SVHN.  F-measure values and percentage of false positives for MNIST and SVHN test datasets are given in Table~\ref{t2}.

\begin{table}
\centering
\begin{tabular}{ |l|l|l| } 
\hline
\textbf{Dataset} & \textbf{MNIST} & \textbf{SVHN} \\
\hline
\textbf{Valid} & MNIST Test & SVHN Test \\
\hline
\textbf{Invalid} & FashionMNIST Test & CIFAR10 Test \\
\hline
\textbf{F-measure} &  0.99 & 0.94 \\
\hline
\textbf{False Positives} & 0.3\% & 2.4\% \\
\hline
\textbf{False Negatives} & 1.42\% & 6.19\% \\
\hline
\end{tabular}
\caption{F-measure and percentage of false positives and false negatives for VAE based input validation model}
\label{t2}
\end{table}

\subsection{Results and Research Questions}\label{results}
In this section, we present results of our experiments we used to answer the research questions.

\textbf{RQ1. Do existing test generation techniques produce invalid inputs?}

We generated test inputs for MNIST and SVHN classifiers using the DeepXplore, DLFuzz and DeepConcolic techniques. The DeepXplore framework supports three types of input transformations: lightening, occlusion and blackout. We generated tests for all three transformations to answer RQ1. 

\begin{figure*}[t!]
  \centering
  \includegraphics[width=\textwidth]{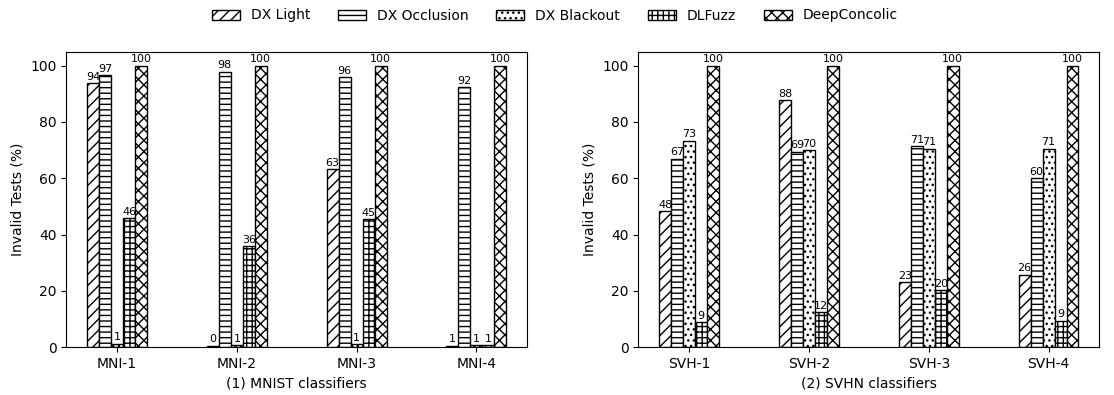}
\includegraphics[width=\textwidth]{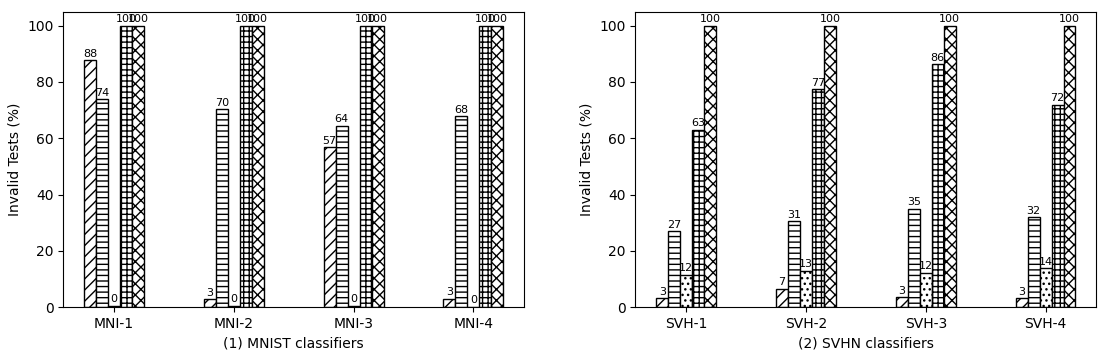}
  \caption{Percentage of invalid test inputs identified by VAE (top pair) and PixelCNN++ (bottom pair) input validation techniques.}
\label{invalid}
\end{figure*}

We randomly sampled 500 seed inputs from each MNIST and SVHN test dataset for DeepXplore and DLFuzz. DeepXplore and DLFuzz use gradient ascent for test generation, and we used the hyperparameters reported in their respective works~\cite{pei2017deepxplore, guo2018dlfuzz} for our study. Similarly, we selected the neuron coverage threshold of 0.25 as it is commonly used in DeepXplore and DLFuzz experiments in their original work.
The DeepConcolic tool uses a single seed input for test generation for neuron coverage, and a timeout of 12 hours is used for test generation in the primary work~\cite{sun2018concolic}. We used the same strategy, and the framework is run with the global optimisation approach. Generated tests are classified as valid or invalid by using the reconstruction probability metric of VAE. The top row of Fig.~\ref{invalid} shows the percentage of invalid test inputs generated by these frameworks for MNIST and SVHN DNN models. 
\begin{table}[t]
\begin{tabular}{|l|l|l|l|l|}
\hline
DNN                      & Testing Technique & Valid (\%) & Invalid (\%) & Total (\%) \\
\hline
\multirow{3}{*}{MNI-1  } & DeepXplore        & 38.5  & \textbf{55.8}    & 55.8  \\
                         & DLFuzz            & 50    & \textbf{50}      & 50    \\
                         & DeepConcolic      & -     & \textbf{55.8}    & 55.8  \\
\hline
\multirow{3}{*}{MNI-2  } & DeepXplore        & 65.5  & \textbf{75}      & 75    \\
                         & DLFuzz            & 71.6  & 70.9    & 71.6  \\
                         & DeepConcolic      & -     & \textbf{58.1}    & 58.1  \\
\hline
\multirow{3}{*}{MNI-3  } & DeepXplore        & 70.9  & \textbf{79.1}    & 79.1  \\
                         & DLFuzz            & 78    & 76.9    & 78    \\
                         & DeepConcolic      & -     & \textbf{64.6}    & 64.6  \\
\hline
\multirow{3}{*}{MNI-4  } & DeepXplore        & 66.6 &  \textbf{73.1}  & 73.3  \\
                         & DLFuzz            & 71.7    &  48.0  &  71.7  \\
                         & DeepConcolic      & -     &  \textbf{63.1}   &  63.1 \\
\hline                         
\end{tabular}
\caption{Neuron Coverage of test inputs generated by DeepXplore, DLFuzz and DeepConcolic for MNIST classifiers}
\label{mnist_nc}
\end{table}

The percentage of tests generated by DeepXplore varies depending on the constraint used. For all the four MNIST classifiers, occlusion constraint produced a high percentage of invalid test inputs i.e., greater than 90\% while blackout constraint generated less than 1\% invalid inputs. The lightening constraint generated 94\% and 63\% invalid inputs for models MNI-1 and MNI-3 and less than 1\% for other two. DLFuzz generated invalid inputs in the range 36\% to 46\% for MNI-1, MNI-2 and MNI-3 classifiers while less than 1\% for MNI-4.

For SVHN classifiers, the occlusion and blackout constraints generated a higher number of invalid tests when compared to lightening constraints on an average. DLFuzz generated invalid inputs are in the range 9\% to 20\% for SVHN classifiers. All the test inputs generated by the DeepConcolic framework for both MNIST and SVHN classifiers are classified as invalid by the VAE model.

\vspace{5pt}
\noindent\fbox{
    \parbox{0.95\linewidth}{
       \textbf{Result for RQ1: All three testing techniques studied produced significant numbers of invalid tests; 42\% on average and ranging from 73-100\% in the worst-case.}
    }
}
\vspace{5pt}

\textbf{RQ2. Existing test generation techniques are guided by test coverage criteria. How do invalid inputs effect test coverage metrics?} 

We measured neuron coverage(NC),  multi-granularity coverage criteria i.e., k-multisection neuron coverage (KMNC), neuron boundary coverage (NBC) and strong neuron activation coverage (SNAC) of both valid and invalid tests generated by the three frameworks. The k-value of 100 is used for measuring KMNC coverage. We also measured the cumulative neuron coverage of valid and invalid test inputs. Results are presented in Tables~\ref{mnist_nc} and \ref{svhn_nc} for neuron coverage metric and Tables~\ref{mnist_kmnc} and \ref{svhn_kmnc} have data for multi-granularity coverage criteria.

Across 8 DNNs, 3 test generation techniques, and 4 coverage criteria, 72\% of the time 
invalid tests achieved coverage greater than or equal to that achieved by valid tests. 
The entries in Tables~\ref{mnist_nc}, \ref{svhn_nc}, \ref{mnist_kmnc} and \ref{svhn_kmnc} corresponding to this insight are highlighted in bold. 
25\% of the time invalid tests outperform valid for coverage, and 25\% of the time invalid coverage boosts overall coverage by more then 10\%.

\vspace{5pt}
\noindent\fbox{%
    \parbox{0.98\linewidth}{%
       \textbf{Result for RQ2: Invalid inputs yield high  coverage for a variety of coverage criterion when compared to valid inputs and they frequently increase coverage beyond that which would be achieved with valid inputs alone.}
    }%
}
\vspace{5pt}

\begin{table}[t]
\begin{tabular}{|l|l|l|l|l|}
\hline
DNN                      & Testing Technique & Valid (\%) & Invalid (\%) & Total (\%) \\
\hline
\multirow{3}{*}{SVH-1} & DeepXplore        & 44.4  & \textbf{44.4}    & 44.4  \\
                         & DLFuzz            & 44.8  & 44.4    & 44.8  \\
                         & DeepConcolic      & -     & \textbf{44.2}    & 44.2  \\
\hline
\multirow{3}{*}{SVH-2} & DeepXplore        & 45.5  & \textbf{45.5}    & 45.5  \\
                         & DLFuzz            & 45.6  & 45.5    & 45.6  \\
                         & DeepConcolic      & -     & \textbf{45.5}    & 45.5  \\
\hline
\multirow{3}{*}{SVH-3} & DeepXplore        & 45.4	 & \textbf{45.4}	   & 45.4  \\
                         & DLFuzz            & 45.7	 & 45.4	   & 45.8  \\
                         & DeepConcolic      & -	 & \textbf{45.2}	   & 45.2  \\
\hline
\multirow{3}{*}{SVH-4} & DeepXplore        & 74.2	 & 74	   & 74.8  \\
                         & DLFuzz            & 75.9	 & 73.3	   & 75.9  \\
                         & DeepConcolic      & -	 & \textbf{72.3}	   & 72.3 \\
\hline                         
\end{tabular}
\caption{Neuron Coverage of test inputs generated by DeepXplore, DLFuzz and DeepConcolic for SVHN classifiers}
\label{svhn_nc}
\end{table}

\begin{table}
\begin{tabular}{|l|l|l|l|l|l|}
\hline
\textbf{DNN}                       & \textbf{\begin{tabular}[c]{@{}l@{}}Testing   \\ Technique\end{tabular}} & \textbf{Coverage} & \textbf{\begin{tabular}[c]{@{}l@{}}Valid   \\ (\%)\end{tabular}} & \textbf{\begin{tabular}[c]{@{}l@{}}Invalid   \\ (\%)\end{tabular}} & \textbf{\begin{tabular}[c]{@{}l@{}}Total   \\ (\%)\end{tabular}} \\
\hline
\multirow{12}{*}{\textbf{MNI-1}} & 
\multirow{4}{*}{\textbf{DeepXplore}}           & \textbf{KMNC}     & 11.3   & \textbf{58}      & 58.8           \\
      &                                        & \textbf{NBC}      & -      & \textbf{1.9}     & 1.9            \\
      &                                        & \textbf{SNAC}     & -      & \textbf{1.9}     & 1.9            \\
      & \multirow{4}{*}{\textbf{DLFuzz}}       & \textbf{KMNC}     & 49.2   & 45.3             & 56.4           \\
      &                                        & \textbf{NBC}      & -     & -                & -              \\
      &                                        & \textbf{SNAC}     & -      & -                & -              \\
      & \multirow{4}{*}{\textbf{DeepConcolic}} & \textbf{KMNC}     & -      & \textbf{8.2}     & 8.2            \\
      &                                        & \textbf{NBC}      & -      & -                & -              \\
      &                                        & \textbf{SNAC}     & -      & -                & -              \\
\hline
\multirow{12}{*}{\textbf{MNI-2}} & 
\multirow{4}{*}{\textbf{DeepXplore}}          & \textbf{KMNC}     & 7.1     & \textbf{62}      & 62.4           \\
     &                                        & \textbf{NBC}      & -       & \textbf{3.4}     & 3.4            \\
     &                                        & \textbf{SNAC}     & -       & \textbf{6.8}     & 6.8            \\
     & \multirow{4}{*}{\textbf{DLFuzz}}       & \textbf{KMNC}     & 49.7    & 40.2             & 54.2           \\
     &                                        & \textbf{NBC}      & -       & -                & -              \\
     &                                        & \textbf{SNAC}     & -       & -                & -              \\
     & \multirow{4}{*}{\textbf{DeepConcolic}} & \textbf{KMNC}     & -       & \textbf{11.2}    & 11.2           \\
     &                                        & \textbf{NBC}      & -       & \textbf{2.4}     & 2.4            \\
     &                                        & \textbf{SNAC}     & -       & \textbf{3.4}     & 3.4            \\
\hline
\multirow{12}{*}{\textbf{MNI-3}} & 
\multirow{4}{*}{\textbf{DeepXplore}}         & \textbf{KMNC}     & 12.5     & \textbf{59.2}    & 59.9           \\
    &                                        & \textbf{NBC}      & 0.2      & \textbf{3.4}     & 3.5            \\
    &                                        & \textbf{SNAC}     & 0.4      & \textbf{6.7}     & 7.1            \\
    & \multirow{4}{*}{\textbf{DLFuzz}}       & \textbf{KMNC}     & 45.8     & 41.6             & 52.2           \\
    &                                        & \textbf{NBC}      & -        & 0.2              & 0.2            \\
    &                                        & \textbf{SNAC}     & -        & -                & -              \\
    & \multirow{4}{*}{\textbf{DeepConcolic}} & \textbf{KMNC}     & -        & \textbf{14.8}    & 14.8           \\
    &                                        & \textbf{NBC}      & -       & \textbf{1.1}     & 1.1           \\
    &                                        & \textbf{SNAC}     & -        & \textbf{2.2}     & 2.2            \\
\hline
\multirow{12}{*}{\textbf{MNI-4}} 
& \multirow{4}{*}{\textbf{DeepXplore}}       & \textbf{KMNC}     &  18.7   & \textbf{56.6}      &    57.5       \\
    &                                        & \textbf{NBC}      &   -     & \textbf{1.8}       &    1.8       \\
    &                                        & \textbf{SNAC}     &   -     & \textbf{2.4}       &    2.4       \\
    & \multirow{4}{*}{\textbf{DLFuzz}}       & \textbf{KMNC}     &  47.5   & 1.6                &     47.5      \\
    &                                        & \textbf{NBC}      & 0.4     & -                  &  0.4         \\
    &                                        & \textbf{SNAC}     &  0.5    & -                  &   0.5        \\
    & \multirow{4}{*}{\textbf{DeepConcolic}} & \textbf{KMNC}     &  -      & \textbf{27.9}      &  27.9         \\
    &                                        & \textbf{NBC}      &  -      & \textbf{3.4}       &     3.4      \\
    &                                        & \textbf{SNAC}     &  -      & \textbf{4.3}       &   4.3        \\                                   
\hline
\end{tabular}
\caption{Multi-granularity neuron coverage of test inputs generated by DeepXplore, DLFuzz and DeepConcolic for MNIST classifiers}
\label{mnist_kmnc}
\end{table}

\begin{table}
\begin{tabular}{|l|l|l|l|l|l|}
\hline
\textbf{DNN}                       & \textbf{\begin{tabular}[c]{@{}l@{}}Testing   \\ Technique\end{tabular}} & \textbf{Coverage} & \textbf{\begin{tabular}[c]{@{}l@{}}Valid   \\ (\%)\end{tabular}} & \textbf{\begin{tabular}[c]{@{}l@{}}Invalid   \\ (\%)\end{tabular}} & \textbf{\begin{tabular}[c]{@{}l@{}}Total   \\ (\%)\end{tabular}} \\
\hline
\multirow{12}{*}{\textbf{SVH-1}} & 
\multirow{4}{*}{\textbf{DeepXplore}}          & \textbf{KMNC}     & 30.7           & \textbf{42.4}    & 46.8           \\
    &                                         & \textbf{NBC}      & 1.6            & \textbf{5.2}     & 6.6            \\
    &                                         & \textbf{SNAC}     & 0.7            & \textbf{9.1}     & 9.3            \\
    & \multirow{4}{*}{\textbf{DLFuzz}}        & \textbf{KMNC}     & 55             & 38.8             & 57.9           \\
    &                                         & \textbf{NBC}      & 2.1            & 0.4              & 2.3            \\
    &                                         & \textbf{SNAC}     & 3.3            & 0.7              & 3.6            \\
    & \multirow{4}{*}{\textbf{DeepConcolic}}  & \textbf{KMNC}     & -              & \textbf{17}      & 17             \\
    &                                         & \textbf{NBC}      & -              & \textbf{1}       & 1              \\
    &                                         & \textbf{SNAC}     & -              & \textbf{2}       & 2            \\
\hline
\multirow{12}{*}{\textbf{SVH-2}} & 
\multirow{4}{*}{\textbf{DeepXplore}}         & \textbf{KMNC}     & 35.2           & \textbf{45.4}    & 50.1           \\
   &                                         & \textbf{NBC}      & 2.3            & 0.7              & 2.6            \\
   &                                         & \textbf{SNAC}     & 1.3            & 1.2              & 1.9            \\
   & \multirow{4}{*}{\textbf{DLFuzz}}        & \textbf{KMNC}     & 58             & 43.5             & 61.3           \\
   &                                         & \textbf{NBC}      & 0.6            & \textbf{1.5}     & 1.8            \\
   &                                         & \textbf{SNAC}     & 1              & \textbf{1.5}     & 2              \\
   & \multirow{4}{*}{\textbf{DeepConcolic}}  & \textbf{KMNC}     & -              & \textbf{22.5}    & 22.5           \\
   &                                         & \textbf{NBC}      & -              & \textbf{0.5}     & 0.5            \\
   &                                         & \textbf{SNAC}     & -              & \textbf{0.8}     & 0.8            \\
\hline
\multirow{12}{*}{\textbf{SVH-3}} & 
\multirow{4}{*}{\textbf{DeepXplore}}         & \textbf{KMNC}     & 32.2           & \textbf{44.9}    & 48.9           \\
   &                                         & \textbf{NBC}      & 2.4            & 0.7              & 2.8            \\
   &                                         & \textbf{SNAC}     & 1.4            & 1.2              & 2.2            \\
   & \multirow{4}{*}{\textbf{DLFuzz}}        & \textbf{KMNC}     & 52.9           & 51               & 60.2           \\
   &                                         & \textbf{NBC}      & 0.5            & \textbf{1.6}     & 1.9            \\
   &                                         & \textbf{SNAC}     & 0.8            & \textbf{2.4}     & 2.8            \\
   & \multirow{4}{*}{\textbf{DeepConcolic}}  & \textbf{KMNC}     & -              & \textbf{20.2}    & 20.2           \\
   &                                         & \textbf{NBC}      & -              & \textbf{5.4}     & 5.4            \\
   &                                         & \textbf{SNAC}     & -              & \textbf{4.5}     & 4.5          \\
\hline
\multirow{12}{*}{\textbf{SVH-4}} & 
\multirow{4}{*}{\textbf{DeepXplore}}        & \textbf{KMNC}      & 24.7          & \textbf{40.4}     &   41.8   \\
    &                                        & \textbf{NBC}      & 1             & \textbf{4}        &     4.1 \\
    &                                        & \textbf{SNAC}     & 1.8           & \textbf{5.5}      &    5.7       \\
    & \multirow{4}{*}{\textbf{DLFuzz}}       & \textbf{KMNC}     & 41.4          & 29.5              &   43.4      \\
    &                                        & \textbf{NBC}      & 1.7           & 1                 &   1.8        \\
    &                                        & \textbf{SNAC}     & 2.7           & 1.8               &  2.9 \\
    & \multirow{4}{*}{\textbf{DeepConcolic}} & \textbf{KMNC}     & -             & \textbf{15.2}     &    15.2       \\
    &                                        & \textbf{NBC}      & -             & \textbf{1.1}      &   1.1        \\
    &                                        & \textbf{SNAC}     & -             & \textbf{2.0}      &    2.0       \\     
\hline
\end{tabular}
\caption{Multi-granularity neuron coverage of test inputs generated by DeepXplore, DLFuzz and DeepConcolic for SVHN classifiers}
\label{svhn_kmnc}
\end{table}


\textbf{RQ3. VAE based input validation can be incorporated into the test generation techniques. How effective is this technique in generating valid inputs and what is the overhead?}

To answer this question, we generated test inputs by using VAE based input validation along with a gradient ascent based test generation technique as described in Algorithm~\ref{alg1}. We selected DeepXplore as the  baseline test generation technique and density estimated by VAE is incorporated as a goal into its objective to formulate a joint optimization. Result of a joint optimization is sensitive to the weights of different goals used in the objective function. To address this, we fixed the weights of the goals of the baseline's objective and performed a sweep over a range of density weights to find the best configuration. We used gradient ascent to generate test inputs for MNIST and SVHN models. We randomly identified 200 seed inputs from each of the two datasets and used the same seed set and gradient ascent parameters, i.e., step size and maximum iterations for baseline and our technique. The experiments are repeated three times and average results are presented in this section.

We measured the number of valid tests generated along with their neuron coverage for our technique and the baseline to demonstrate the effectiveness of our technique. The validity of the inputs is measured with respect to the OOD detection algorithm used, i.e., the VAE in this case. Our technique generates only valid test inputs. Since baseline generates both valid and invalid test inputs, we added the input validation module to the baseline to capture only the valid test inputs.

\begin{figure*}
  \centering
  \includegraphics[width=\textwidth]{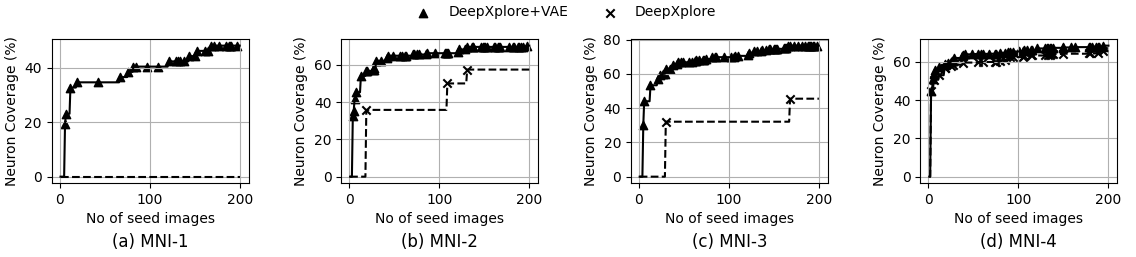}
  \caption{Neuron Coverage of valid inputs generated by DeepXplore and DeepXplore extended with VAE for MNIST models}
\label{mnist_occl_gen}
\end{figure*}

\begin{figure*}
  \centering
  \includegraphics[width=\textwidth]{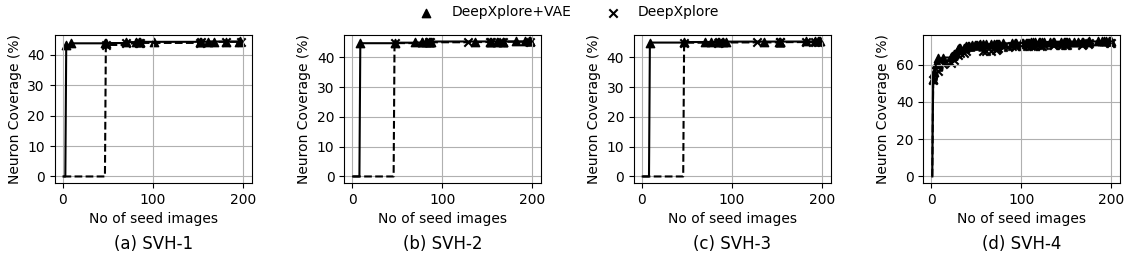}
  \caption{Neuron Coverage of valid inputs generated by DeepXplore and DeepXplore extended with VAE for SVHN models}
\label{svhn_occl_gen}
\end{figure*}

\begin{figure}
  \centering
  \includegraphics[width=\linewidth]{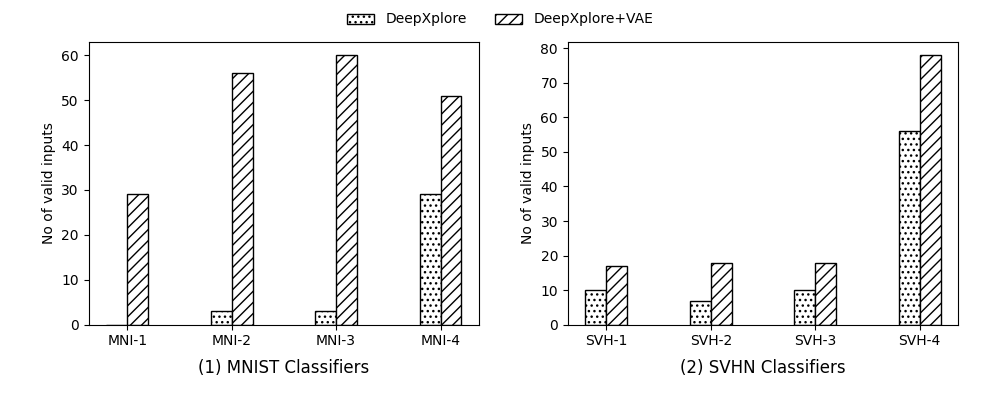}
  \caption{Number of valid inputs generated by DeepXplore and DeepXplore extended with VAE for MNIST and SVHN models}
\label{mnist_svhn_occl}
\end{figure}

Neuron coverage achieved by the baseline technique and our technique are presented in Figures~\ref{mnist_occl_gen} and ~\ref{svhn_occl_gen} for MNIST and SVHN classifiers respectively. The plots show the coverage over a range of 200 seed inputs. Our technique achieved neuron coverage greater than or equivalent to that of DeepXplore baseline for all the 8 DNN models. For the scenarios where baseline is able achieve neuron coverage comparable to ours, our technique outperformed the baseline in terms of the number of valid inputs generated. Fig.~\ref{mnist_svhn_occl} contains a comparison of the number of valid inputs generated by the baseline and our technique for MNIST and SVHN classifiers. The total valid inputs generated by our technique for the MNIST models are 5.6 times the valid inputs generated by the baseline. For SVHN dataset, our technique generated 1.6 times more valid inputs when compared to the baseline. Hence, having VAE in the test objective guides gradient ascent effectively in searching for valid inputs.

Table ~\ref{test_gen_perf} shows the performance data of DeepXplore+VAE and DeepXplore algorithms for 200 seed inputs. Every iteration of these algorithms has two components, 1) gradient ascent, and 2) input validation. For each seed input, gradient ascent is performed until it finds a valid test input or for a maximum of 30 iterations whichever happens first. Input validation is performed only when the differential oracle fails the generated test input in that iteration. In all the cases, DeepXplore+VAE ran for fewer iterations and input validations when compared to the baseline. For the scenarios where the difference between DeepXplore+VAE and baseline's number of iterations and input validations is high, DeepXplore+VAE is faster because the baseline is spending more time on generating invalid inputs which are then rejected by the input validation module. When this difference is small, baseline has better overall run-time, but DeepXplore+VAE generates more valid inputs and has lower cost per valid input when compared to the baseline. We note that due to DeepXplore+VAE's improved effectiveness in generating valid tests it improves on the baseline's "time to produce a valid test" reducing it from 4.7 to 1.7 minutes, on average measured across three runs.

\vspace{5pt}
\noindent\fbox{
    \parbox{0.95\linewidth}{
       \textbf{Result for RQ3: Incorporating a VAE into test generation eliminates the generation of invalid test inputs, significantly increases the generation of valid inputs, reduces the time to generate valid tests, and increases coverage achieved on generated valid tests.}
    }
}
\vspace{5pt}

\begin{table*}
\centering
\begin{tabular}{|l|l|l|l|l|l|l|l|l|l|l|} 
\hline
\multicolumn{1}{|c|}{\multirow{2}{*}{DNN}} & \multicolumn{4}{c|}{DeepXplore+VAE}                                                                                                                          & \multicolumn{4}{c|}{DeepXplore}                                                                                                                              & \multicolumn{1}{c|}{\multirow{2}{*}{\begin{tabular}[c]{@{}l@{}l@{}l@{}}Iterations \\ (DeepXplore+VAE \\ - DeepXplore)\end{tabular} }} & \multicolumn{1}{c|}{\multirow{2}{*}{\begin{tabular}[c]{@{}l@{}l@{}l@{}}Input validations \\ (DeepXplore+VAE \\ - DeepXplore)\end{tabular}}}  \\ 
\cline{2-9}
\multicolumn{1}{|c|}{}                     & \multicolumn{1}{c|}{\begin{tabular}[c]{@{}l@{}}Run-time   \\ in mins\end{tabular}} & \multicolumn{1}{c|}{\begin{tabular}[c]{@{}l@{}}Valid   \\ Inputs\end{tabular}} & \multicolumn{1}{c|}{Iterations} & \multicolumn{1}{c|}{\begin{tabular}[c]{@{}l@{}}Input \\ validations\end{tabular}} & \multicolumn{1}{c|}{\begin{tabular}[c]{@{}l@{}}Run-time \\ in mins\end{tabular}} & \multicolumn{1}{c|}{\begin{tabular}[c]{@{}l@{}}Valid \\ inputs\end{tabular}} & \multicolumn{1}{c|}{Iterations} & \multicolumn{1}{c|}{\begin{tabular}[c]{@{}l@{}}Input \\ validations\end{tabular}} & \multicolumn{1}{c|}{}                                                                          & \multicolumn{1}{c|}{}                                                                                  \\ 
\hline
MNI-1                                      & 96.74                                     & 29                                  & 5413                            & 882                                      & 103.82                                    & 1                                   & 5972                            & 1832                                     & -559                                                                                           & -950                                                                                                   \\ 
\hline
MNI-2                                      & 73.5                                      & 54                                  & 4910                            & 413                                      & 103                                       & 3                                   & 5913                            & 1812                                     & -1003                                                                                          & -1399                                                                                                  \\ 
\hline
MNI-3                                      & 60.39                                     & 56                                  & 4863                            & 200                                      & 96.66                                     & 3                                   & 5917                            & 1587                                     & -1054                                                                                          & -1387                                                                                                  \\ 
\hline
MNI-4                                      & 54.97                                     & 52                                  & 4736                            & 52                                       & 46.57                                     & 29                                  & 5199                            & 375                                      & -463                                                                                           & -323                                                                                                   \\ 
\hline
SVH-1                                      & 97.12                                     & 17                                  & 5637                            & 27                                       & 64.7                                      & 12                                  & 5737                            & 47                                       & -100                                                                                           & -20                                                                                                    \\ 
\hline
SVH-2                                      & 97.96                                     & 20                                  & 5578                            & 28                                       & 66.83                                     & 9                                   & 5798                            & 60                                       & -220                                                                                           & -32                                                                                                    \\ 
\hline
SVH-3                                      & 90.34                                     & 21                                  & 5539                            & 29                                       & 69.83                                     & 11                                  & 5703                            & 80                                       & -164                                                                                           & -51                                                                                                    \\ 
\hline
SVH-4                                      & 143.81                                    & 83                                  & 4126                            & 219                                      & 130.57                                    & 53                                  & 4547                            & 446                                      & -421                                                                                           & -227                                                                                                   \\
\hline
\end{tabular}
\caption{Run-time analysis of test generation algorithms of DeepXplore+VAE and DeepXplore for MNIST and SVHN classifiers}
\label{test_gen_perf}
\end{table*}

\textbf{RQ4. Is the determination of invalid inputs sensitive to the generative model used?}

To answer RQ4, we use a PixelCNN++ based input validation technique. PixelCNN++ is an autoregressive deep generative model~\cite{salimans2017pixelcnn++}. The advantage of using this model for out-of-distribution detection is that the model outputs the probability density explicitly. We trained PixelCNN++ models for MNIST and SVHN datasets. For each dataset, we find the threshold for identifying invalid inputs by using an invalid dataset and F-measure analysis similar to VAE based detection technique described in Section~\ref{threshold}. The F-measure, precision and recall of the selected thresholds for both the datasets are presented in Table~\ref{pcnn_threshold}.

The percentage of test inputs generated by DeepXplore, DLFuzz and DeepConcolic for the MNIST and SVHN classification models that are classified as invalid by PixelCNN++ based input classifier are presented on the bottom row of Fig.~\ref{invalid}.  PixelCNN++ for the MNIST models, classified a high percentage of test inputs generated by DeepXplore's light and occlusion constraints as invalid and classified all test inputs as valid for blackout constraint. For the SVHN classifiers, occlusion and blackout constraints result in higher number of invalid inputs when compared to the  light constraint. 

The PixelCNN++ classified all test inputs generated by DLFuzz as invalid for MNIST models and more than 60\% test inputs as invalid for SVHN models. All inputs generated by DeepConcolic are identified as invalid for both the models.

The results follow the same trend as observed by VAE based classifier. 
However the percentage of test inputs classified as invalid by PixelCNN++ is less when compared to that of VAE for DeepXplore generated tests. For DLFuzz, the PixelCNN++ approach resulted in more invalid tests when compared to the VAE based classifier. Both the VAE and PixelCNN++ based techniques classified all test inputs generated by DeepConcolic as invalid.

\vspace{5pt}
\noindent\fbox{%
    \parbox{0.98\linewidth}{%
       \textbf{Result for RQ4: Test generators are judged to produce invalid tests with different OOD techniques, but the number of invalid tests is sensitive to the deep generative model architecture used.}
    }%
}
\vspace{5pt}

\begin{table}
\centering
\begin{tabular}{ |l|l|l| } 
\hline
\textbf{Dataset} & \textbf{MNIST} & \textbf{SVHN} \\
\hline
\textbf{Valid} & MNIST Test & SVHN Test \\
\hline
\textbf{Invalid} & FashionMNIST Test & CIFAR10 Test \\
\hline
\textbf{F-measure} &  0.99 & 0.92 \\
\hline
\textbf{False Positives} & 0.14\% & 2\%  \\
\hline
\textbf{False Negatives} & 0.56\% & 10.66\% \\
\hline
\end{tabular}
\caption{F-measure and percentage of false positives and false negatives for PixelCNN++ based input validation model}
\label{pcnn_threshold}
\end{table}

\section{Threats to Validity}\label{threats}
We designed our study to provide a degree of generalizability by
spanning all of the algorithmic families of DNN test generation approaches that have been developed to date, as well as 2 datasets, 8 models, 4 coverage criteria, and 2 approaches to out-of-distribution detection.  Moreover, the datasets and models that we have chosen are those that have been used in prior research -- which was both a convenience choice and a means of promoting comparison among methods, e.g., against baselines.  Despite these measures, our findings may be dependent on these choices.  

Further study, especially with additional OOD techniques, beyond VAE and PixelCNN++, is warranted to understand the generalizability of our findings as relates to the rate at which invalid inputs are generated and the degree of coverage achieved by those inputs.  Our study on adapting test generation with OOD is more limited using a single model, a VAE, and a single test generation approach, DeepXplore which is a representative of the class of optimization-based test generation approaches.   It is not a simple matter to extend this study to other families of test generation methods, but that will be necessary to understand the extent to which the benefit of integrating OOD methods with DNN test generation techniques broadly generalizes.

We ran all of our experiments multiple times and cross-checked them with prior work, e.g., that we achieved the same level of coverage for baseline techniques as was reported in prior work. We took these measures to assure the quality of the data reported here and we made the code available in github for transparency and replicability.

\section{Conclusions}\label{conclusions}
This paper demonstrates that existing DNN test generation and test coverage techniques do not consider the 
valid input space, which can have several deleterious effects.  It can lead DNN test  methods to generate large numbers of invalid inputs -- those that lie off the training distribution as judged by state-of-the-art techniques -- thereby reducing the efficiency of the test generation process and, even worse, producing large 
numbers of tests that might be rejected as invalid during fault 
\newpage
\noindent 
triage processes.   It can lead test coverage techniques to value invalid tests inappropriately by achieving or improving on coverage from valid tests -- this has the potential to bias test generation results.

\begin{wrapfigure}{r}{0.4\columnwidth}
\begin{center}
\begin{lstlisting}
(*$\DNN_{\mathit{defensive}}$*)((*$\vec{i}$*)) {
  if (!OOD((*$\vec{i}$*)))
    return (*$\DNN$*)((*$\vec{i}$*))
  else
    return error((*$\vec{i}$*))
}
\end{lstlisting}
\end{center}
\caption{Defensive DNN}
\label{fig:defensivednn}
\end{wrapfigure}

We demonstrate that existing out of distribution detection techniques can be coupled with
test generation algorithms to address this problem.  In this work, we focused on VAE-based
OOD detection and incorporating such models into optimization-based test generation.  Our study shows this to be effective in significantly boosting the number of valid test inputs 
generated and in eliminating invalid tests. While promising, more work is needed to explore
the potential for other OOD models to inform test generation and to incorporate such
models into constraint-based and fuzzing test generators.  

Finally, we plan to explore how the well-understood concept
of defensive programming for traditional programs, as sketched in Fig.~\ref{fig:code}, can
be adapted to DNNs.   Fig.~\ref{fig:defensivednn} sketches a possibility suggested
by the findings in this paper, where the role of input validation is played by an OOD
detector.  In such an architecture, testing of $\DNN$ should be restricted to inputs that
are not out of distribution, but testing of the
OOD itself must be conducted over a broader input space as is the case with prior work
on input validation testing~\cite{sinha2000analysis,zhang2014amplifying,goffi2016automatic}.  With such an architecture, DNN test suites that achieve higher coverage of OOD and $\DNN$
are \textit{better}, thereby reestablishing the long held intuitions about test coverage for traditional software.

\section*{Acknowledgements}
This material is based in part upon work supported by 
National Science Foundation awards 1900676 and 2019239.

\clearpage
\bibliographystyle{IEEEtran.bst}
\bibliography{IEEEfull, main}

\begin{thebibliography}{10}
\providecommand{\url}[1]{#1}
\csname url@samestyle\endcsname
\providecommand{\newblock}{\relax}
\providecommand{\bibinfo}[2]{#2}
\providecommand{\BIBentrySTDinterwordspacing}{\spaceskip=0pt\relax}
\providecommand{\BIBentryALTinterwordstretchfactor}{4}
\providecommand{\BIBentryALTinterwordspacing}{\spaceskip=\fontdimen2\font plus
\BIBentryALTinterwordstretchfactor\fontdimen3\font minus
  \fontdimen4\font\relax}
\providecommand{\BIBforeignlanguage}[2]{{%
\expandafter\ifx\csname l@#1\endcsname\relax
\typeout{** WARNING: IEEEtran.bst: No hyphenation pattern has been}%
\typeout{** loaded for the language `#1'. Using the pattern for}%
\typeout{** the default language instead.}%
\else
\language=\csname l@#1\endcsname
\fi
#2}}
\providecommand{\BIBdecl}{\relax}
\BIBdecl

\bibitem{end2endautonomous2017}
\BIBentryALTinterwordspacing
M.~Bojarski, D.~D. Testa, D.~Dworakowski, B.~Firner, B.~Flepp, P.~Goyal, L.~D.
  Jackel, M.~Monfort, U.~Muller, J.~Zhang, X.~Zhang, J.~Zhao, and K.~Zieba,
  ``End to end learning for self-driving cars,'' \emph{CoRR}, vol.
  abs/1604.07316, 2016. [Online]. Available:
  \url{http://arxiv.org/abs/1604.07316}
\BIBentrySTDinterwordspacing

\bibitem{pendleton2017perception}
S.~Pendleton, H.~Andersen, X.~Du, X.~Shen, M.~Meghjani, Y.~Eng, D.~Rus, and
  M.~Ang, ``Perception, planning, control, and coordination for autonomous
  vehicles,'' \emph{Machines}, vol.~5, no.~1, p.~6, 2017.

\bibitem{TrailNet:IROS2017}
N.~{Smolyanskiy}, A.~{Kamenev}, J.~{Smith}, and S.~{Birchfield}, ``Toward
  low-flying autonomous mav trail navigation using deep neural networks for
  environmental awareness,'' in \emph{2017 IEEE/RSJ International Conference on
  Intelligent Robots and Systems (IROS)}, Sep. 2017, pp. 4241--4247.

\bibitem{loquercio-etal:RAL:2018:dronet}
A.~Loquercio, A.~I. Maqueda, C.~R.~D. Blanco, and D.~Scaramuzza, ``Dronet:
  Learning to fly by driving,'' \emph{{IEEE} Robotics and Automation Letters},
  2018.

\bibitem{pei2017deepxplore}
K.~Pei, Y.~Cao, J.~Yang, and S.~Jana, ``Deepxplore: Automated whitebox testing
  of deep learning systems,'' in \emph{proceedings of the 26th Symposium on
  Operating Systems Principles}, 2017, pp. 1--18.

\bibitem{ma2018deepgauge}
L.~Ma, F.~Juefei-Xu, F.~Zhang, J.~Sun, M.~Xue, B.~Li, C.~Chen, T.~Su, L.~Li,
  Y.~Liu \emph{et~al.}, ``Deepgauge: Multi-granularity testing criteria for
  deep learning systems,'' in \emph{Proceedings of the 33rd ACM/IEEE
  International Conference on Automated Software Engineering}, 2018, pp.
  120--131.

\bibitem{sun2018testingmcdc}
Y.~Sun, X.~Huang, D.~Kroening, J.~Sharp, M.~Hill, and R.~Ashmore, ``Testing
  deep neural networks,'' \emph{arXiv preprint arXiv:1803.04792}, 2018.

\bibitem{DBLP:journals/jss/XieHMKXC11}
\BIBentryALTinterwordspacing
X.~Xie, J.~W.~K. Ho, C.~Murphy, G.~E. Kaiser, B.~Xu, and T.~Y. Chen, ``Testing
  and validating machine learning classifiers by metamorphic testing,''
  \emph{J. Syst. Softw.}, vol.~84, no.~4, pp. 544--558, 2011. [Online].
  Available: \url{https://doi.org/10.1016/j.jss.2010.11.920}
\BIBentrySTDinterwordspacing

\bibitem{tian2018deeptest}
Y.~Tian, K.~Pei, S.~Jana, and B.~Ray, ``Deeptest: Automated testing of
  deep-neural-network-driven autonomous cars,'' in \emph{Proceedings of the
  40th international conference on software engineering}, 2018, pp. 303--314.

\bibitem{huang2020survey}
X.~Huang, D.~Kroening, W.~Ruan, J.~Sharp, Y.~Sun, E.~Thamo, M.~Wu, and X.~Yi,
  ``A survey of safety and trustworthiness of deep neural networks:
  Verification, testing, adversarial attack and defence, and
  interpretability,'' \emph{Computer Science Review}, vol.~37, p. 100270, 2020.

\bibitem{hayes2006input}
J.~H. Hayes and J.~Offutt, ``Input validation analysis and testing,''
  \emph{Empirical Software Engineering}, vol.~11, no.~4, pp. 493--522, 2006.

\bibitem{li2010perturbation}
N.~Li, T.~Xie, M.~Jin, and C.~Liu, ``Perturbation-based user-input-validation
  testing of web applications,'' \emph{Journal of Systems and Software},
  vol.~83, no.~11, pp. 2263--2274, 2010.

\bibitem{liu2009covering}
H.~Liu and H.~B.~K. Tan, ``Covering code behavior on input validation in
  functional testing,'' \emph{Information and Software Technology}, vol.~51,
  no.~2, pp. 546--553, 2009.

\bibitem{taneja2010mitv}
K.~Taneja, N.~Li, M.~R. Marri, T.~Xie, and N.~Tillmann, ``Mitv:
  multiple-implementation testing of user-input validators for web
  applications,'' in \emph{Proceedings of the IEEE/ACM international conference
  on Automated software engineering}, 2010, pp. 131--134.

\bibitem{sinha2000analysis}
S.~Sinha and M.~J. Harrold, ``Analysis and testing of programs with exception
  handling constructs,'' \emph{IEEE Transactions on Software Engineering},
  vol.~26, no.~9, pp. 849--871, 2000.

\bibitem{zhang2014amplifying}
P.~Zhang and S.~Elbaum, ``Amplifying tests to validate exception handling code:
  An extended study in the mobile application domain,'' \emph{ACM Transactions
  on Software Engineering and Methodology (TOSEM)}, vol.~23, no.~4, pp. 1--28,
  2014.

\bibitem{goffi2016automatic}
A.~Goffi, A.~Gorla, M.~D. Ernst, and M.~Pezz{\`e}, ``Automatic generation of
  oracles for exceptional behaviors,'' in \emph{Proceedings of the 25th
  International Symposium on Software Testing and Analysis}, 2016, pp.
  213--224.

\bibitem{sun2018concolic}
Y.~Sun, M.~Wu, W.~Ruan, X.~Huang, M.~Kwiatkowska, and D.~Kroening, ``Concolic
  testing for deep neural networks,'' in \emph{Proceedings of the 33rd ACM/IEEE
  International Conference on Automated Software Engineering}, 2018, pp.
  109--119.

\bibitem{guo2018dlfuzz}
J.~Guo, Y.~Jiang, Y.~Zhao, Q.~Chen, and J.~Sun, ``Dlfuzz: Differential fuzzing
  testing of deep learning systems,'' in \emph{Proceedings of the 2018 26th ACM
  Joint Meeting on European Software Engineering Conference and Symposium on
  the Foundations of Software Engineering}, 2018, pp. 739--743.

\bibitem{an2015variational}
J.~An and S.~Cho, ``Variational autoencoder based anomaly detection using
  reconstruction probability,'' \emph{Special Lecture on IE}, vol.~2, no.~1,
  2015.

\bibitem{xu2018unsupervised}
H.~Xu, W.~Chen, N.~Zhao, Z.~Li, J.~Bu, Z.~Li, Y.~Liu, Y.~Zhao, D.~Pei, Y.~Feng
  \emph{et~al.}, ``Unsupervised anomaly detection via variational auto-encoder
  for seasonal kpis in web applications,'' in \emph{Proceedings of the 2018
  World Wide Web Conference}, 2018, pp. 187--196.

\bibitem{zenati2018efficient}
H.~Zenati, C.~S. Foo, B.~Lecouat, G.~Manek, and V.~R. Chandrasekhar,
  ``Efficient gan-based anomaly detection,'' \emph{arXiv preprint
  arXiv:1802.06222}, 2018.

\bibitem{chalapathy2019deep}
R.~Chalapathy and S.~Chawla, ``Deep learning for anomaly detection: A survey,''
  \emph{arXiv preprint arXiv:1901.03407}, 2019.

\bibitem{lecun1998mnist}
Y.~LeCun, ``The mnist database of handwritten digits,'' \emph{http://yann.
  lecun. com/exdb/mnist/}, 1998.

\bibitem{netzer2011reading}
Y.~Netzer, T.~Wang, A.~Coates, A.~Bissacco, B.~Wu, and A.~Y. Ng, ``Reading
  digits in natural images with unsupervised feature learning,'' 2011.

\bibitem{zhang2020machine}
J.~M. Zhang, M.~Harman, L.~Ma, and Y.~Liu, ``Machine learning testing: Survey,
  landscapes and horizons,'' \emph{IEEE Transactions on Software Engineering},
  2020.

\bibitem{goodfellow2014explaining}
I.~J. Goodfellow, J.~Shlens, and C.~Szegedy, ``Explaining and harnessing
  adversarial examples,'' \emph{arXiv preprint arXiv:1412.6572}, 2014.

\bibitem{kurakin2016adversarial}
A.~Kurakin, I.~Goodfellow, and S.~Bengio, ``Adversarial examples in the
  physical world,'' \emph{arXiv preprint arXiv:1607.02533}, 2016.

\bibitem{carlini2017towards}
N.~Carlini and D.~Wagner, ``Towards evaluating the robustness of neural
  networks,'' in \emph{2017 ieee symposium on security and privacy (sp)}.\hskip
  1em plus 0.5em minus 0.4em\relax IEEE, 2017, pp. 39--57.

\bibitem{stuctural_coverage}
E.~J. Weyuker, ``The evaluation of program-based software test data adequacy
  criteria,'' \emph{Communications of the ACM}, vol.~31, no.~6, pp. 668--675,
  1988.

\bibitem{ma2018combinatorial}
L.~Ma, F.~Zhang, M.~Xue, B.~Li, Y.~Liu, J.~Zhao, and Y.~Wang, ``Combinatorial
  testing for deep learning systems,'' \emph{arXiv preprint arXiv:1806.07723},
  2018.

\bibitem{odena2019tensorfuzz}
A.~Odena, C.~Olsson, D.~Andersen, and I.~Goodfellow, ``Tensorfuzz: Debugging
  neural networks with coverage-guided fuzzing,'' in \emph{International
  Conference on Machine Learning}, 2019, pp. 4901--4911.

\bibitem{hendrycks2018deep}
D.~Hendrycks, M.~Mazeika, and T.~Dietterich, ``Deep anomaly detection with
  outlier exposure,'' \emph{arXiv preprint arXiv:1812.04606}, 2018.

\bibitem{nguyen2015deep}
A.~Nguyen, J.~Yosinski, and J.~Clune, ``Deep neural networks are easily fooled:
  High confidence predictions for unrecognizable images,'' in \emph{Proceedings
  of the IEEE conference on computer vision and pattern recognition}, 2015, pp.
  427--436.

\bibitem{kingma2013auto}
D.~P. Kingma and M.~Welling, ``Auto-encoding variational bayes,'' \emph{arXiv
  preprint arXiv:1312.6114}, 2013.

\bibitem{goodfellow2014generative}
I.~Goodfellow, J.~Pouget-Abadie, M.~Mirza, B.~Xu, D.~Warde-Farley, S.~Ozair,
  A.~Courville, and Y.~Bengio, ``Generative adversarial nets,'' in
  \emph{Advances in neural information processing systems}, 2014, pp.
  2672--2680.

\bibitem{van2016conditional}
A.~Van~den Oord, N.~Kalchbrenner, L.~Espeholt, O.~Vinyals, A.~Graves
  \emph{et~al.}, ``Conditional image generation with pixelcnn decoders,'' in
  \emph{Advances in neural information processing systems}, 2016, pp.
  4790--4798.

\bibitem{salimans2017pixelcnn++}
T.~Salimans, A.~Karpathy, X.~Chen, and D.~P. Kingma, ``Pixelcnn++: Improving
  the pixelcnn with discretized logistic mixture likelihood and other
  modifications,'' \emph{arXiv preprint arXiv:1701.05517}, 2017.

\bibitem{lecun1998gradient}
Y.~LeCun, L.~Bottou, Y.~Bengio, and P.~Haffner, ``Gradient-based learning
  applied to document recognition,'' \emph{Proceedings of the IEEE}, vol.~86,
  no.~11, pp. 2278--2324, 1998.

\bibitem{springenberg2014striving}
J.~T. Springenberg, A.~Dosovitskiy, T.~Brox, and M.~Riedmiller, ``Striving for
  simplicity: The all convolutional net,'' \emph{arXiv preprint
  arXiv:1412.6806}, 2014.

\bibitem{simonyan2014vggnet}
K.~Simonyan and A.~Zisserman, ``Very deep convolutional networks for
  large-scale image recognition,'' \emph{arXiv preprint arXiv:1409.1556}, 2014.

\bibitem{nalisnick2018deep}
E.~Nalisnick, A.~Matsukawa, Y.~W. Teh, D.~Gorur, and B.~Lakshminarayanan, ``Do
  deep generative models know what they don't know?'' \emph{arXiv preprint
  arXiv:1810.09136}, 2018.

\bibitem{krizhevsky2009learning}
A.~Krizhevsky, G.~Hinton \emph{et~al.}, ``Learning multiple layers of features
  from tiny images,'' 2009.

\bibitem{ren2019likelihood}
J.~Ren, P.~J. Liu, E.~Fertig, J.~Snoek, R.~Poplin, M.~Depristo, J.~Dillon, and
  B.~Lakshminarayanan, ``Likelihood ratios for out-of-distribution detection,''
  in \emph{Advances in Neural Information Processing Systems}, 2019, pp.
  14\,707--14\,718.

\bibitem{xiao2017fashionmnist}
H.~Xiao, K.~Rasul, and R.~Vollgraf. (2017) Fashion-mnist: a novel image dataset
  for benchmarking machine learning algorithms.

\bibitem{rosca2018distribution}
M.~Rosca, B.~Lakshminarayanan, and S.~Mohamed, ``Distribution matching in
  variational inference,'' \emph{arXiv preprint arXiv:1802.06847}, 2018.

\end{thebibliography}

\end{document}